# Producing High Concentrations of Hydrogen in Palladium via Electrochemical Insertion from Aqueous and Solid Electrolytes

*Jesse D. Benck, Ariel Jackson, David Young, Daniel Rettenwander, Yet-Ming Chiang\**

Department of Materials Science and Engineering, Massachusetts Institute of Technology.

77 Massachusetts Ave, Cambridge, MA 02139.

\*Author to whom correspondence should be addressed

# Abstract

Metal hydrides are critical materials in numerous technologies including hydrogen storage, gas separation, and electrocatalysis. Here, using Pd-H as a model metal hydride, we perform electrochemical insertion studies of hydrogen via liquid and solid state electrolytes at 1 atm ambient pressure, and achieve H:Pd ratios near unity, the theoretical solubility limit. We show that the compositions achieved result from a dynamic balance between the rate of hydrogen insertion and evolution from the Pd lattice, the combined kinetics of which are sufficiently rapid that *operando* experiments are necessary to characterize instantaneous $PdH_x$ composition. We use simultaneous electrochemical insertion and X-ray diffraction measurements, combined with a new calibration of lattice parameter versus hydrogen concentration, to enable accurate quantification of the composition of electrochemically synthesized $PdH_x$. Furthermore, we show that the achievable hydrogen concentration is severely limited by electrochemomechanical damage to the palladium and/or substrate. The understanding embodied in these results helps to establish new design rules for achieving high hydrogen concentrations in metal hydrides.

# Introduction

Hydrogen absorption in metals is critical to a number of technological applications including hydrogen storage,[1-6] superconductivity,[7-9] gas separation,[10-13] and hydrogen embrittlement.[14,15] In electrocatalysis, changing the binding energies of surface-adsorbed intermediates by straining or alloying materials is widely employed as a design strategy to improve reaction rates.[16-18] Control of metal-hydrogen composition (e.g., $x$ in $MH_x$) influences adsorbate binding energies that modulate the activity and selectivity of important electrocatalytic reactions such as the hydrogen evolution reaction (HER) and $CO_2$ reduction reaction. Understanding how and when electrochemistry can substitute for the high pressures previously used to create $MH_x$ with H:M ratios far exceeding the equilibrium values at ambient pressure is of fundamental and practical interest. Electrochemomechanical stresses analogous to those created by H insertion and extraction from the metal lattice are important as a material degradation mechanism in lithium ion batteries[19-21] and solid oxide fuel cells.[22] Hydrogen embrittlement is a well-known metallurgical failure mode[23,24] that has found new significance in hydrogen-based energy systems, for example, as a process central to the longevity of hydrogen transportation pipelines.[25]

Pd is widely used to study metal-hydrogen phenomena due to the large dynamic range of $PdH_x$ composition at accessible temperatures and pressures, the high H diffusion coefficient in Pd, and the facile dissociation of molecular $H_2$ at the surface of Pd. These properties enable readily tunable bulk $PdH_x$ compositions with H:Pd ratio between 0 and 0.7. However, as we show, achieving higher H loading (H:Pd ratio > 0.7) in a controlled and reproducible manner, which is desirable for both basic studies and some of the above-mentioned applications, is not trivial.

The most widely studied means of forming $PdH_x$ from Pd are the physical application of $H_2$ gas pressure, and the electrochemical insertion of hydrogen. The relationship between $H_2$ gas pressure, $\beta$-$PdH_x$ composition, and temperature is well-documented,[26-28] but for every 0.1 increase in H:Pd ratio above 0.7 an increase of about 1.5 orders of magnitude of $H_2$ pressure is needed, making highly loaded $PdH_x$ challenging to synthesize outside specialized laboratory facilities. Electrochemistry, on the other hand, can in theory achieve the same thermodynamic activities of hydrogen via the convenient and straightforward application of modest electrochemical potentials, and at ambient pressure and temperature. For example, achieving a H:Pd ratio of 1 at room temperature requires $H_2$ pressures in the GPa range,[29] but the equivalent thermodynamic Nernst potential is only ~100 mV at room temperature. When combined with kinetic overpotentials present in typical electrochemical cells, the total cell voltage remains on the order of 1-10 V, which is easily accessible in most experimental designs.

However, electrochemical H insertion suffers from a gap in understanding of the quantitative relationship between current, potential, and hydrogen concentration. In our view this is largely due to the difficulty of accurately measuring $PdH_x$ composition during electrochemistry. A variety of measurement techniques have been employed to quantify this relationship, but each comes with its own limitations. One method for quantifying H:Pd ratio is to outgas the sample and measure the evolved gas after the electrochemical experiment is complete.[27,28,30] However, this method correlates composition with the ensemble electrochemical profile, rather than the instantaneous potential or current, does not account for residual hydrogen left behind, and may not capture transients, given that the equilibrium pressure of a $PdH_x$ sample at high hydrogen composition (H:Pd > 0.7) is several thousand atmospheres.[31]

*Operando* measurements, in principle, can avoid such measurement errors and the poor time resolution of *ex situ* techniques. A number of previous studies have attempted *operando* measurements of the PdH$_x$ composition, but have encountered other sources of error. Electrochemical quartz crystal microbalance (EQCM) measurements[32] require non-trivial corrections to frequency shifts caused by the H insertion-induced strain. Electrochemical oxidation (coulometry) of the absorbed H[33] is subject to both (1) H:Pd ratio overestimation from oxidation of coevolved hydrogen gas remaining from the insertion process and (2) H:Pd ratio underestimation due to gas desorption during switching from reducing to oxidizing potentials. The widely-used resistance ratio method[34] aims to take advantage of the dependence of Pd electrical resistivity on hydrogen concentration, which exhibits a maximum at x~0.7.[35] However, this technique is complicated by the existence of two compositions that correspond to the same resistance ratio, any variations in temperature during measurement, irreversible resistivity changes due to stress-induced microstructure evolution in the Pd alloy, and electrical shunt pathways through the electrolyte. These and other complications of the resistivity method have been described in detail by Zhang et al.[36] Amongst previously used methods, *operando* measurements of PdH$_x$ structure such as X-ray diffraction (XRD) or extended X-ray absorption fine structure (EXAFS) probably have the least uncertainty. Typically, synchrotron radiation has been used to measure the H:Pd ratio via XRD measurements of lattice parameter[37-40] or EXAFS measurement of interatomic distances.[41] Still, proper calibration of structure dimensions to H:Pd ratio is needed.

In this study, we developed a novel methodology, and instrumentation, for electrochemically inserting hydrogen into Pd from both liquid and solid electrolytes. Using a three-electrode method, we report for the first time the dependence of the PdH$_x$ composition at high H:Pd ratios (> 0.7) on the cathode potential alone (as opposed to the full cell voltage, which depends on the anode performance). For accurate quantification of composition, we designed an apparatus that enables *operando* structure measurements during H insertion via powder X-ray diffractometry. Higher experimental throughput is afforded than with a synchrotron-based apparatus, allowing reproducibility and sample-to-sample variations to be assessed. We identify calibration errors in prior work and develop an improved calibration of the dependence of *β*-PdH$_x$ lattice parameter on H:Pd ratio with quantified uncertainty, thereby enabling both more accurate and more precise determination of the PdH$_x$ composition than in previous studies.

In pursuit of high, controllable hydrogen loading levels, we investigated cell configurations that accommodate the use of three classes of electrolytes: An aqueous electrolyte, a proton-conducting solid polymer electrolyte (Nafion™), and a proton-conducting ceramic solid electrolyte (yttria and ceria doped barium zirconate). Systematic study over a wide range of experimental conditions was conducted, with particular focus on effects of the type of electrode-electrolyte interface, the cathode thickness, and temperature. Our results reveal that the use of thin cathodes, low temperatures (~25 °C), and electrode-electrolyte structures resistant to electrochemomechanical damage maximize the achievable H:Pd ratio.

## Experimental Section

### Sample Configurations

Figure 1 A, D, and G show the experimental configuration for aqueous electrolyte cell tests. Pd cathodes of two types were used in the aqueous electrolyte cells. One consisted of free-standing 25 μm thick Pd foils (Beantown Chemical), while the other consisted of 50 nm thick Pd thin films sputtered onto a 127 μm thick Kapton film substrate, with a Kapton shadow mask defining the electrode geometry. All thin films used in this work were sputtered using a Q300TD sputterer

(Quorum Technologies) at a deposition rate of approximately 10 nm/min. A 4 nm Cr adhesion layer was sputtered between the Kapton and the Pd film. Both cathodes had surface areas of 0.8 - 1.2 cm$^2$. The edges of each Pd foil and thin film cathode were taped down onto the Kapton substrate with Kapton adhesive to aid in positioning the foils and to protect the adhesion layer of the thin films. The anode was a 25 μm thick Pt foil (Beantown Chemical). The cell was assembled with a gap between the cathode and anode, that was subsequently flooded with aqueous electrolyte consisting of 0.05 M $H_2SO_4$ (Sigma-Aldrich) in ultrapure water (18.2 MΩ cm, VWR). The reference electrode was Ag/AgCl in saturated KCl (Radiometer Analytical). Hydrogen gas bubbles nucleate in the electrolyte-filled gap of this cell.

The solid electrolyte cells have a different sample configuration from the liquid electrolyte cell, as shown in Figure 1, in that the Pd sample is attached directly to the solid electrolyte. The Nafion™-based electrochemical cells were fabricated using 2 cm diameter, 127 μm thick disks of Nafion™ 117 electrolyte (Sigma-Aldrich). Figure 1 B, E, and H show the experimental configuration. Prior to use, the Nafion™ was pre-treated by boiling in 0.5 M $H_2SO_4$ for at least 1 hr, then rinsed thoroughly in water and dried in air. Pd film cathodes of 50 nm thickness and 1.13 cm$^2$ surface area were sputtered directly onto the electrolyte through a Kapton mask. A 10 - 15 nm thick Pt anode and reversible hydrogen electrode (RHE) reference electrode were sputtered onto the opposite side of each Nafion™ disk through a Kapton mask. No adhesion layer was used between the Nafion™ and cathode or anode. To make electrical contact to the external circuit, current collectors were cut from porous carbon paper into the shape of the anode, cathode, and reference electrode, and attached to the Pd and Pt electrodes, providing uniform electrical contact while still enabling gas access. These carbon current collectors are X-ray transparent and do not interfere with the *operando* structure measurement.

The BaZr$_{0.8}$Ce$_{0.1}$Y$_{0.1}$O$_3$ (BZCY)-based electrochemical cells (Fig. 1 F and I) were fabricated using 2 cm diameter, 1 mm thick disks of BCZY (CoorsTek). Prior to use, the BZCY disks were polished on both sides to an RMS roughness of ~30 nm using an automatic polishing wheel (Allied Multiprep) using alumina and diamond slurries. Pd cathodes of 200 nm thickness and 1.13 cm$^2$ surface area were sputtered directly onto the electrolyte through a Kapton mask. A Pt anode of 10 - 15 nm thickness and a Pt RHE reference electrode were sputtered onto the opposite side of each BZCY disk through a Kapton mask. A 4 nm Cr adhesion layer was sputtered onto the BZCY before the Pt was deposited.

*Electrochemical Cell Design*

Two custom apparatuses were designed and fabricated for use with aqueous and solid electrolytes, respectively, to enable XRD simultaneously with electrochemical measurements. The aqueous electrolyte test cell, shown in Figure 1A, was 3D printed using a Mojo 3D printer (Stratasys) from acrylonitrile butadiene styrene (ABS) plastic. The Kapton-Pd cathode assembly was adhered to the top of the apparatus using silicone adhesive, to create a liquid-tight electrolyte well. The cathode is centered above the electrolyte well, in the X-ray beam spot, and also extends beyond the silicone seal to provide a dry electrical lead. Kapton was chosen for its low X-ray attenuation and good chemical resistance to the electrolyte. Electrical connections were made to the Pd cathode leads outside of the electrolyte well using Cu tape. The Pt foil anode was attached with epoxy adhesive to a glass slide positioned at the bottom of the electrolyte well. The electrolyte well was filled with the aqueous electrolyte through the vent of the apparatus, following which the reference electrode was inserted into the vent. To maintain a flooded cell even during vigorous gas bubble generation, the top of the cell was designed such that the surface slopes upwards towards the vent tube (Fig. 1A).

The solid electrolyte test cell was designed for XRD concurrently with electrochemical operation in $H_2$ gas environment at temperatures from room temperature up to 745 °C (Figure 1B). This apparatus features a stainless steel housing, gas-tight seals, gas supply tubes, platinum electrical leads to the electrochemical cell, a heater for operation at elevated temperatures, an S-type thermocouple temperature probe (Omega), Be foil X-ray windows for low X-ray attenuation, and a removable optically transparent quartz viewport on the top of the device. A Nafion™ or BZCY sample assembly was inserted into the device through the open top flange and secured onto the heater block with ceramic screws. Electrical connections were made to the cathode, anode, and reference electrode using Pt current collectors (Figure 1C). The entire cell chamber was then closed by sealing the quartz viewport onto the top CF flange with a Cu gasket. A detailed description of the apparatus appears elsewhere.[42]

*Operando* **XRD**

Both the aqueous and solid-electrolyte cells were designed to be mounted in a SmartLab multipurpose X-ray diffractometer (Rigaku) for *operando* XRD experiments. Except where otherwise specified, the diffractometer was fitted with parallel beam optics so that the diffraction spectra would not be sensitive to sample height displacement error. Electrical connections were made between the apparatus and a VSP potentiostat (Bio-Logic). In the case of the solid electrolyte apparatus, gas connections were made to provide humidified $H_2$ or Ar gas, and for BZCY samples, the heater was connected to a temperature controller and power source. BZCY samples were heated in Ar until reaching ~200 °C, whereupon $H_2$ was introduced. Both apparatuses for aqueous and solid-electrolyte cells were aligned such that the X-ray beam was incident on the center of the cathode. In the case of the solid electrolyte apparatus, the X-ray beam entered and exited through the Be windows. For the aqueous apparatus, the X-ray beam passed through the Kapton film that held the Pd film or foil.

Full scans from 15 to 90° were used to determine the baseline reflection positions prior to starting electrochemistry. Typically, shorter scans ranging between 36 to 42° were utilized to obtain a time resolution of 3 - 10 min per scan during electrochemistry. Either galvanostatic steps or galvanodynamic sweeps ranging from 0 to 300 mA/cm$^2$ were applied to the electrochemical cells. Three-electrode electrochemical impedance spectroscopy was performed using a frequency range of 100 mHz to 500 kHz and an excitation amplitude of 100 µA.

XRD data were analyzed using HighScore Plus (Panalytical) software. Reflection positions were acquired using Pawley or Rietveld fitting in HighScore Plus. The fits to a set of XRD spectra collected using a representative aqueous electrolyte cell are shown in Supporting Information Video S1. The fitted reflection positions were converted to a lattice parameter, and then to a H:Pd ratio using the calibration curve described in the "Calibration of Lattice Parameter vs H:Pd for PdH$_x$" section below.

*Microstructural Characterization of Pd Cathodes*

Prior to and following experiments, Pd cathode surface morphology was examined using a scanning electron microscope (FEI Helios) operated with an accelerating voltage of 5 kV in secondary electron mode. This instrument was also used to perform energy dispersive X-ray spectroscopy (EDS) with an accelerating voltage of 15 kV. During the XRD measurements, cathodes were also optically monitored using a small USB camera mounted inside the diffractometer chamber.

*Experiments Conducted*

Multiple experiments, under a range of electrodynamic conditions discussed below in the Results section, were conducted on each of the four different electrochemical cell types shown in Table 1. These cell configurations allowed the effects of cathode thickness, electrolyte type, and temperature on the H:Pd ratio to be systematically explored.

**Table 1.** Cathode design and experimental capabilities of electrochemical cell types used.

| Electrochemical Cell Type | Cathode Format | Cathode Thickness | Electrolyte Composition | Temperature Range | Gas Atmosphere |
|---|---|---|---|---|---|
| **Aqueous electrolyte, Pd foil** | Pd foil | 25 μm | 0.05 M $H_2SO_4$ in ultrapure water | 23 to 27 °C | 1 atm air |
| **Aqueous electrolyte, Pd film** | Sputtered thin film | 50 nm | 0.05 M $H_2SO_4$ in ultrapure water | 23 to 27 °C | 1 atm air |
| **Solid polymer electrolyte, Pd film** | Sputtered thin film | 50 nm | Nafion™ 117 | 23 to 80°C | 1 atm humidified $H_2$ |
| **Ceramic electrolyte, Pd film** | Sputtered thin film | 200 nm | $BaZr_{0.8}Ce_{0.1}Y_{0.1}O_3$ (BZCY) | 23 to 750°C | 1 atm humidified $H_2$ |

Views of the experimental apparatuses, schematic designs, and samples appear in Figure 1. Figure 1A is an external view of the cell used for aqueous electrolytes, and Figures 1B and 1C show external and internal views of the apparatus used with Nafion™ and BZCY solid electrolytes. In each case, XRD is conducted on the Pd cathode while the applied voltage and/or current are varied over time. The solid electrolyte cell allows introduction of a humidified $H_2$ gas stream and temperature control over the range 25 - 750 °C. Figures 1D-F show schematic views of the experimental configurations, and Figures 1G-I show photographs of the Pd-electrolyte sample assemblies. In the aqueous electrolyte experiments (Figures 1D and 1G), XRD is conducted on the Pd foil or thin film cathode through the Kapton film to which it is attached, and the Pd cathode is separated from the Pt anode by a 1 cm gap filled with the aqueous electrolyte.

In the Nafion™ electrolyte experiments, (Figures 1E and 1H), the Pd cathode and Pt anode + reference electrode are separated by the 127 μm thick Nafion™ electrolyte. Similarly, in the BZCY electrolyte experiments (Figures 1F and 1I), the Pd cathode is separated from the Pt anode + reference electrode by the BZCY electrolyte, which is 1 mm thick.

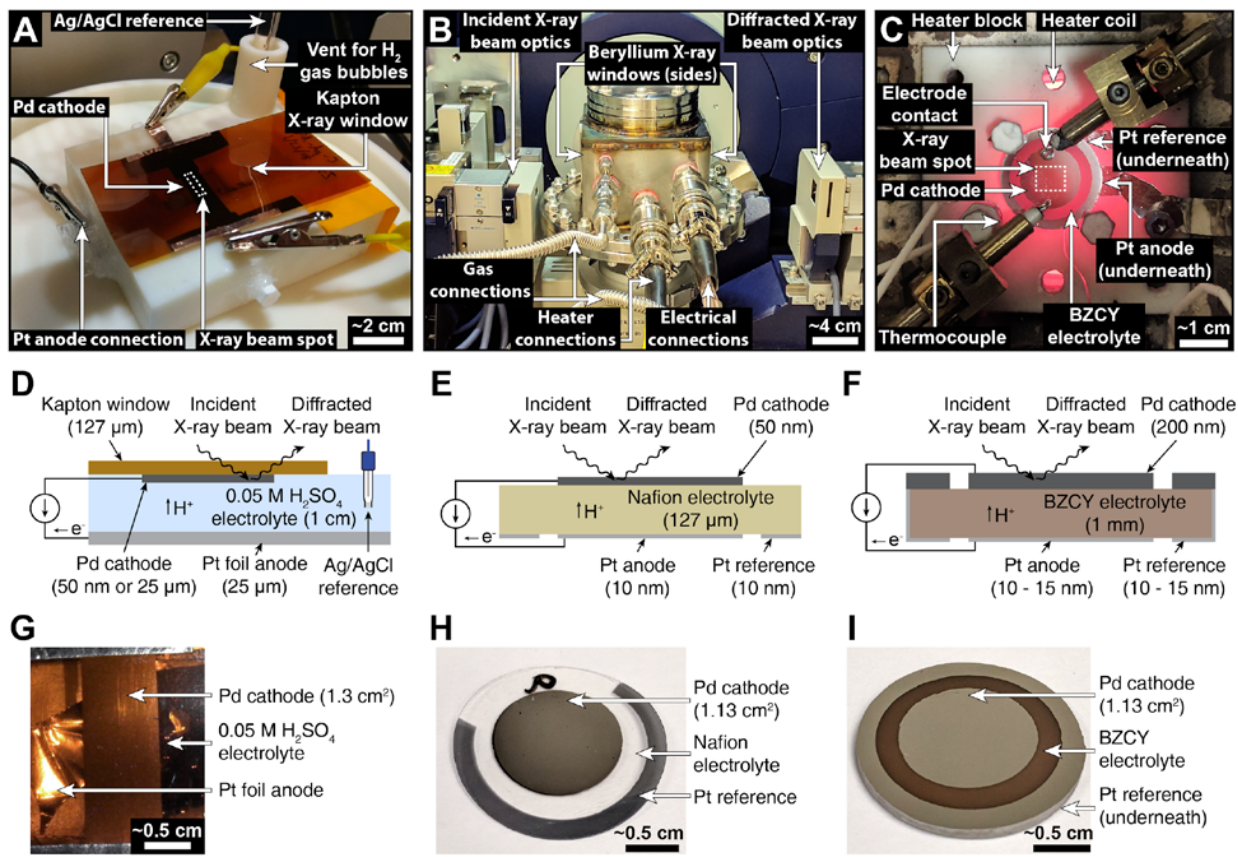

**Figure 1.** Electrochemical cells for *operando* X-ray diffraction (XRD) during electrochemical insertion of hydrogen into Pd foils and thin films. (A) Aqueous electrolyte apparatus and corresponding (D) schematic of experimental configuration and (G) photograph of sample in operating position. (B) Exterior and (C) interior views of the solid electrolyte apparatus. In (C), the experiment is operating at 700 °C. Schematics of the experimental configuration and corresponding photographs of the solid sample assembly are shown in (E, H) for Nafion™ and (F, I) for $BaZr_{0.8}Ce_{0.1}Y_{0.1}O_3$ (BZCY), respectively.

In the aqueous electrolyte experiments, the cathode thickness affects the Pd atom-normalized H flux during electrochemical H insertion, which may influence the maximum H:Pd ratio that can be achieved. A liquid electrolyte is able to form conformal contact with the cathode except as impacted by gas bubble evolution. The use of a solid electrolyte for hydrogen insertion with a Pd-gas phase interface for hydrogen evolution avoids gas bubbles entirely, and is a "dry" experiment that exposes the Pd surface. This format is of interest for interrogation by other analytical methods not possible in a liquid cell. As we show, the extent of adhesion of the Pd to the solid electrolyte dramatically affects hydrogen insertion. Limited work on the use of solid electrolytes appears in the literature.[39,40] It was of interest to us to explore temperature regimes well above those accessible by aqueous or Nafion™ electrolytes, firstly to explore the temperature dependence of hydrogen insertion *per se*, and secondly because a vacancy-rich PdH$_x$ phase with a H:Pd ratio greater than 1 has been reported at high temperatures (700-800 °C) and hydrogen pressures (several GPa).[29] Results presented below are primarily for a temperature of 400 °C, at which there is sufficient proton conductivity to achieve a reasonable cell impedance. The Nafion™ electrolyte cell could also be operated over a temperature range commensurate with sample stability, although results for room temperature are presented here. The aqueous electrolyte experiments were conducted at room temperature.

## Results and Discussion

### *Operando* XRD of PdH$_x$

A typical example of XRD spectra collected during electrochemical H insertion is shown in Figure 2, here for a 50 nm thick Pd cathode in the aqueous electrolyte cell. As shown in Figure 2A, a 15 - 90° scan of the Pd cathode prior to electrochemical H insertion contains four Pd reflections. The (111) reflection has the highest intensity, while the (200), (311), and (222) reflections have lower relative intensity than expected for a randomly oriented polycrystalline film, indicating a <111> preferred orientation. Figure 2B shows 30 spectra collected between 36 - 42° every 3 min before and during electrochemical H insertion. The first four scans at open circuit show the Pd (111) reflection at 40.10°, which corresponds to a Pd lattice parameter of 3.892 A, closely matching the literature value of 3.889 A.[28] Beginning with the fifth scan and ending with the final scan, the cathodic current density was continuously ramped at constant rate from 0 to -0.1 mA/cm$^2$. During this current ramp, H was inserted into the cathode, resulting in the formation of the $β$-PdH$_x$ phase. This phase transformation resulted in a reflection shift to a final position of 38.34°, which corresponds to a lattice parameter of 4.063 A, a 4.4% increase over that of pure Pd. Similar examples of XRD spectra collected from electrochemical cells using Pd foil in aqueous electrolyte, the Nafion™-based solid electrolyte cell, and the BZCY-based solid electrolyte cell, are shown in Supporting Information Figures S3, S4, and S5, respectively.

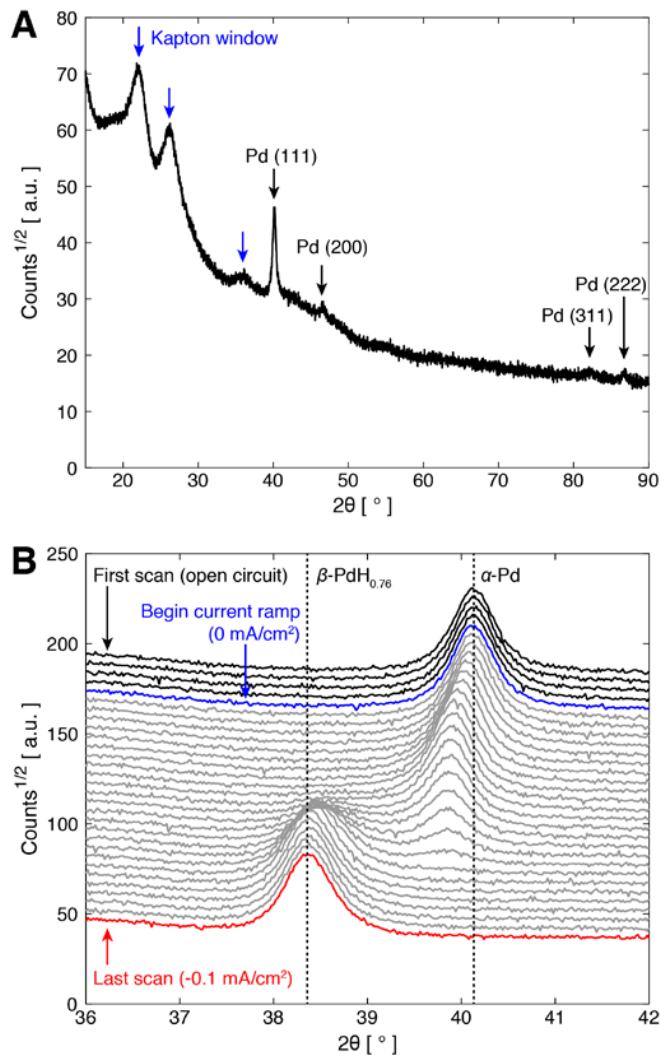

**Figure 2.** (A) XRD spectrum of a representative Pd thin film cathode mounted in the aqueous electrolyte cell prior to electrochemical H insertion. Both $α$-Pd and Kapton reflections were observed, as expected. A $2θ$ scan range between 15° and 90° was used to ascertain the proper alignment of the Pd cathode, as well as to search for impurity phases. (B) XRD spectra around the (111) reflection of $PdH_x$ during electrochemical insertion of H into a Pd thin film cathode in the aqueous electrolyte cell. The spectra are vertically offset for clarity. The leftward shift in the reflection position as current was continuously ramped from 0 to -0.1 mA/cm$^2$ indicates an expansion in the lattice parameter. The discontinuity of the reflection shift, as well as temporary coexistence of two separate reflection, corresponds to a phase transition from $α$-Pd to $β$-PdH$_x$.

## Calibration of Lattice Parameter vs H:Pd for PdH$_x$

To determine the amount of H inserted into these Pd cathodes from the XRD spectra, we developed a new calibration of the lattice parameter vs H:Pd ratio in $\beta$-PdH$_x$ (Vegard's relationship). Although previous correlations exist, they have been created from sparse data sets and were not rigorously corrected for the effects of thermal expansion and/or pressure compression.[26,27,31,38] Furthermore, we discovered that one of the widely used calibration curves, from Manchester and coworkers[31] (their eq 7) mistakenly uses data collected at -196 °C[26] to obtain a room-temperature calibration curve. Studies using this correlation will overestimate the H:Pd ratio from $\beta$-PdH$_x$ lattice parameter measurements.

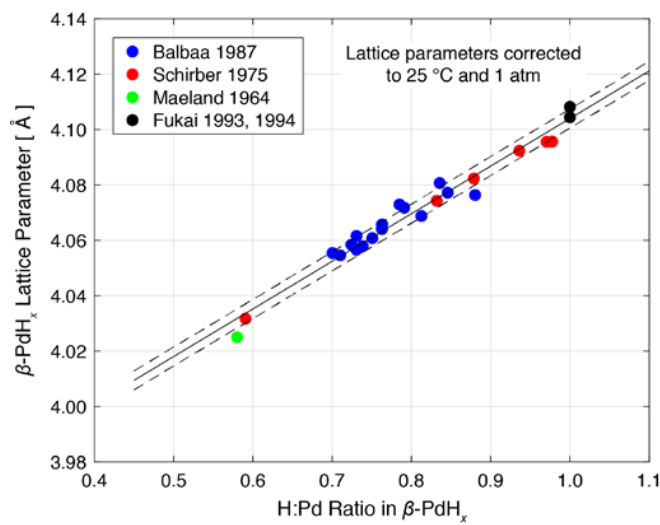

**Figure 3.** Calibration of PdH$_x$ lattice parameter vs H:Pd ratio using data from references.[26-29,43] All lattice parameter values were corrected to standard temperature and pressure (25 °C, 1 atm) using literature values for thermal expansion[44,45] and bulk modulus.[46] The solid line is a linear regression fit, while the dotted lines represent offsets of one standard deviation.

We developed an improved calibration using data from five references in which the $\beta$-PdH$_x$ lattice parameter was measured via XRD, and the H:Pd ratio of the same sample was independently determined via outgassing,[26,27] weighing,[28] or assuming a fixed composition $\beta$-PdH$_x$ during a two phase transition.[29,43] The first three sets of measurements[26-28] were performed at liquid nitrogen temperatures of -196 °C and under pressures up to 4,000 atm. The other two measurements by Fukai[29,43] were done at temperatures around 700-800 °C and extremely high pressures up to about 50,000 atm. We corrected the measured lattice parameters for thermal expansion/contraction and pressure compression back to 25 °C and 1 atm using published coefficients of thermal expansion[44,45] and bulk modulus.[46] Details of these corrections are described in Supporting Information Section S2. The resulting data are displayed in Figure 3, which shows that the $\beta$-PdH$_x$ lattice parameter corrected to 25 °C and 1 atm varies linearly with H:Pd ratio in $\beta$-PdH$_x$. The solid line represents the following relationship between lattice parameter, $a$, and H:Pd ratio derived from a linear least squares regression fit to the data:

$$a\ [\text{A}] = 3.9321 + 0.1719 \times \text{H:Pd} \qquad (1)$$

The dashed lines represent $\pm\ \sigma_a = 0.0034$ A, one standard deviation in the measured lattice parameter around the fit line, which corresponds to an uncertainty in the H:Pd ratio of $\pm\ \sigma_{H:Pd} = 0.020$.

**Hydrogen Insertion Results**

The results of representative H:Pd measurements from each sample type are shown as time-series data in Figures 4 and 5. In each case the current density was varied over time in a stepwise manner, and the corresponding cathode potential and lattice parameter were simultaneously measured. The corresponding H:Pd ratio in the $\beta$-PdH$_x$ phase was then calculated from the lattice parameter using eq 1. These data reveal how cathode composition varies in each sample as a function of the current density, cathode potential, and time.

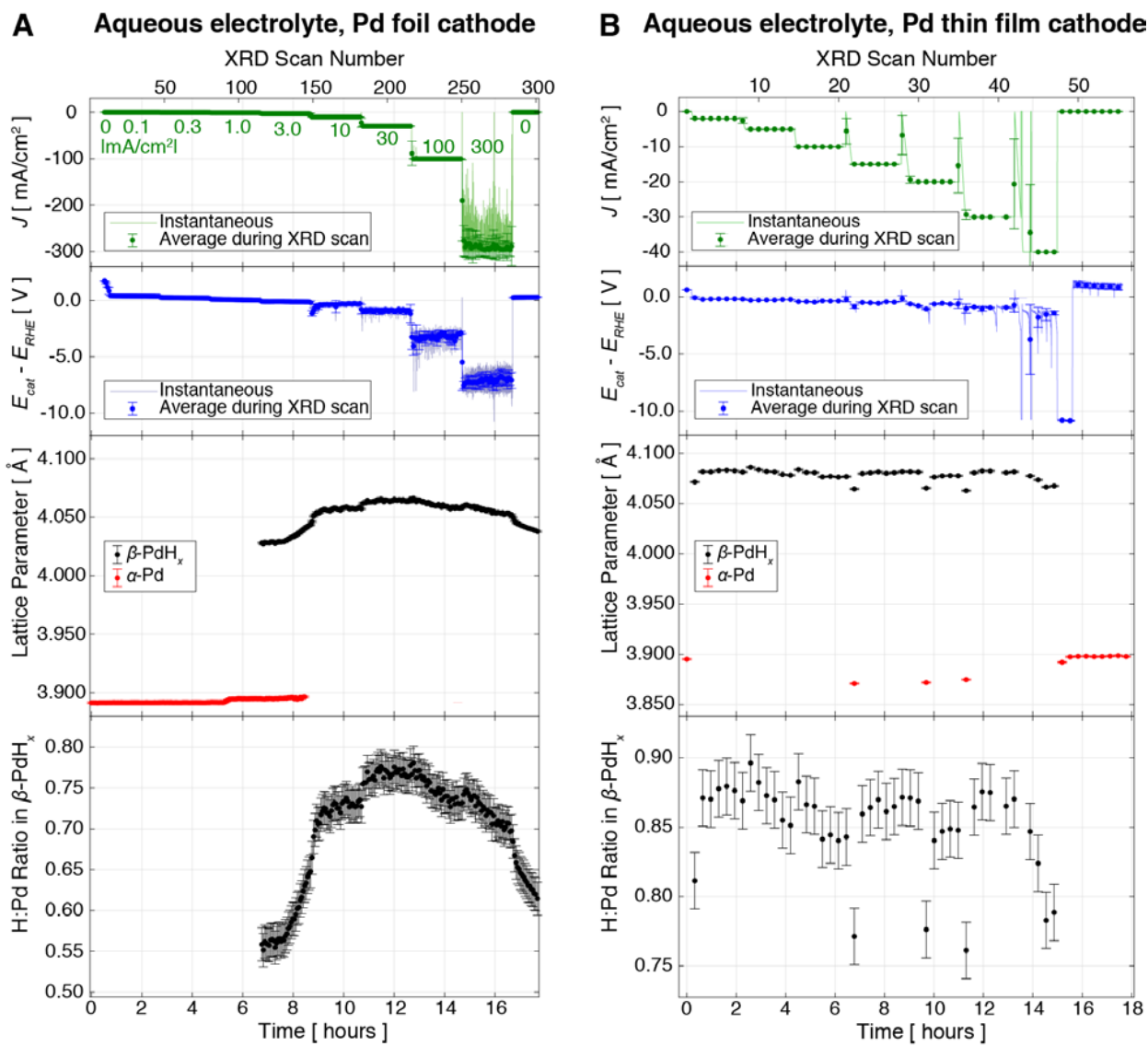

**Figure 4.** Representative results for aqueous electrochemical insertion of hydrogen into (A) Pd foil and (B) Pd thin film, at room temperature. The top two panels display applied current density ($J$) vs time, and the corresponding cathode potential ($E_{cat}$) vs. reversible hydrogen electrode potential ($E_{RHE}$). The third panel from the top shows the PdH$_x$ lattice parameter vs time. The bottom panel shows the corresponding H:Pd ratio for the $\beta$-PdH$_x$ phase obtained using the calibration curve in Figure 3.

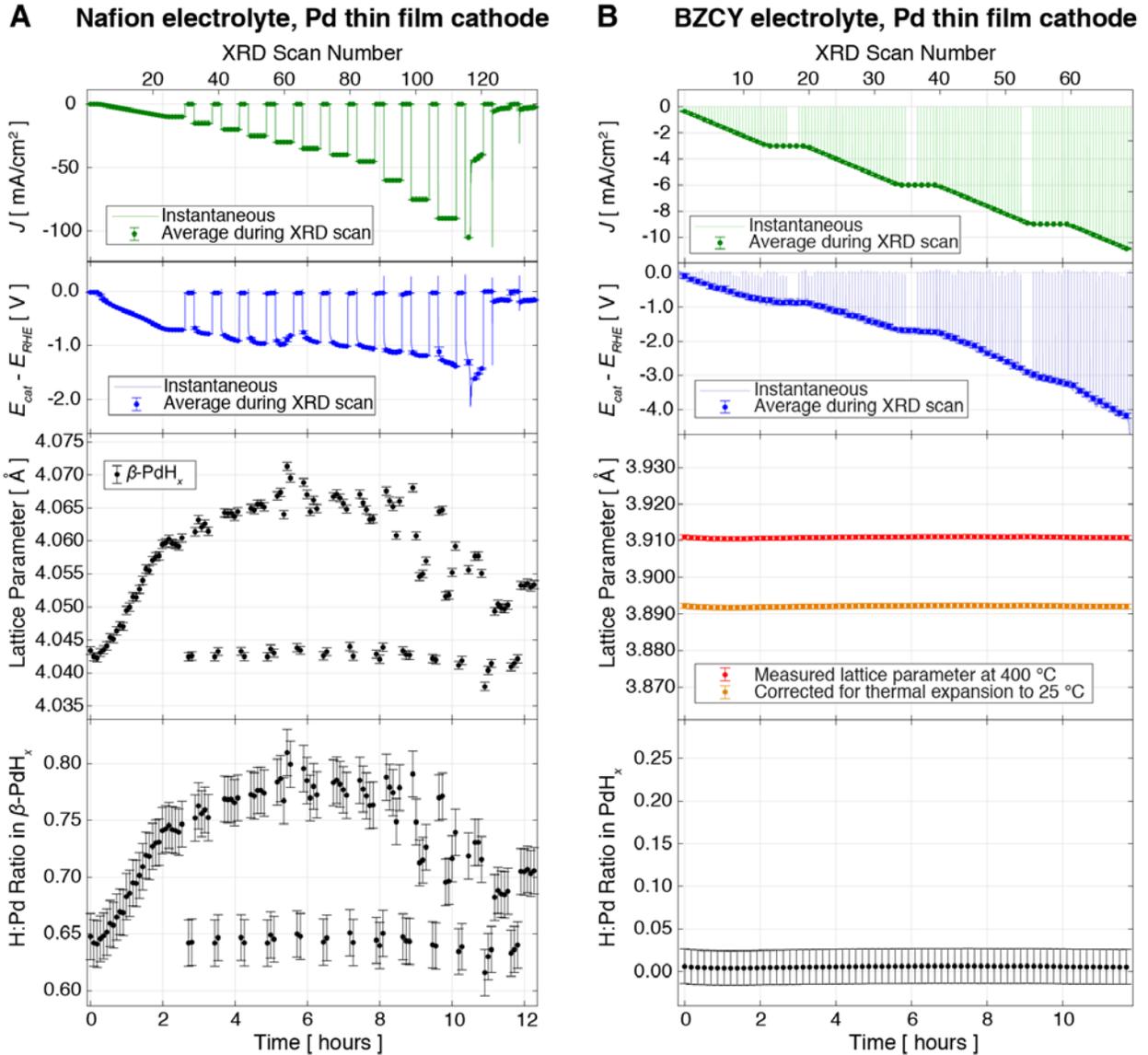

**Figure 5.** Representative results for electrochemical insertion of hydrogen into Pd thin film cathodes on (A) Nafion™ and (B) BZCY solid electrolytes, at room temperature and 400 °C, respectively. A humidified $H_2$ gas environment is used. The top two panels display applied current density ($J$) vs time, and the corresponding cathode potential ($E_{cat}$) vs. reversible hydrogen electrode potential ($E_{RHE}$). The third panel from the top shows the PdH$_x$ lattice parameter vs time. For the Pd cathode on BZCY, the lattice parameter at 400 °C measured has been corrected to its room temperature values using thermal expansion data. The bottom left panel shows the H:Pd ratio for the $\beta$-PdH$_x$ phase on Nafion™, derived from the calibration curve in Figure 3. The bottom right panel shows the H:Pd ratio for the Pd film on BZCY, obtained using a calibration for $\alpha$-Pd at low H:Pd ratio from Eastman et al.[47]

The top two panels in Figures 4 and 5 show the instantaneous current density and cathode potential (relative to RHE) as lines, and the time-averaged current density and cathode potential during each 3 - 20 min XRD scan as points, plotted against time. The error bars for the data points in each instance represents one standard deviation of the measured values during the 3 - 20 min XRD scan interval. Additional electrochemical data including the anode potential for each of these electrochemical cells is displayed in Supporting Information Figure S12. The third panel from the top in Figures 4 and 5 displays the $\alpha$-Pd and/or $\beta$-PdH$_x$ lattice parameters, determined by fitting each of the XRD spectra as described in the Experimental Methods section against time. The lattice parameter measurement error is ± 0.00061 A, determined by calculating the standard deviation of fitted lattice parameters from 39 spectra of a single sample under identical open circuit conditions. Finally, the bottom panel in Figures 4 and 5 show the H:Pd ratio in the $\beta$-PdH$_x$ phase calculated from the lattice parameter using eq 1 for the aqueous foil, aqueous thin film, and Nafion™ thin film samples. The error bars are ± 0.020, corresponding to one standard deviation uncertainty from the H:Pd ratio/lattice parameter correlation in Figure 3. Uncertainty in the lattice parameter measurement is negligible compared to the uncertainty from the H:Pd ratio/lattice parameter correlation. Because the H:Pd ratio observed in the BZCY thin films was low enough to be in the $\alpha$-Pd range, we used eq 1 from Eastman et al.[47] to determine the H:Pd ratio for this sample.

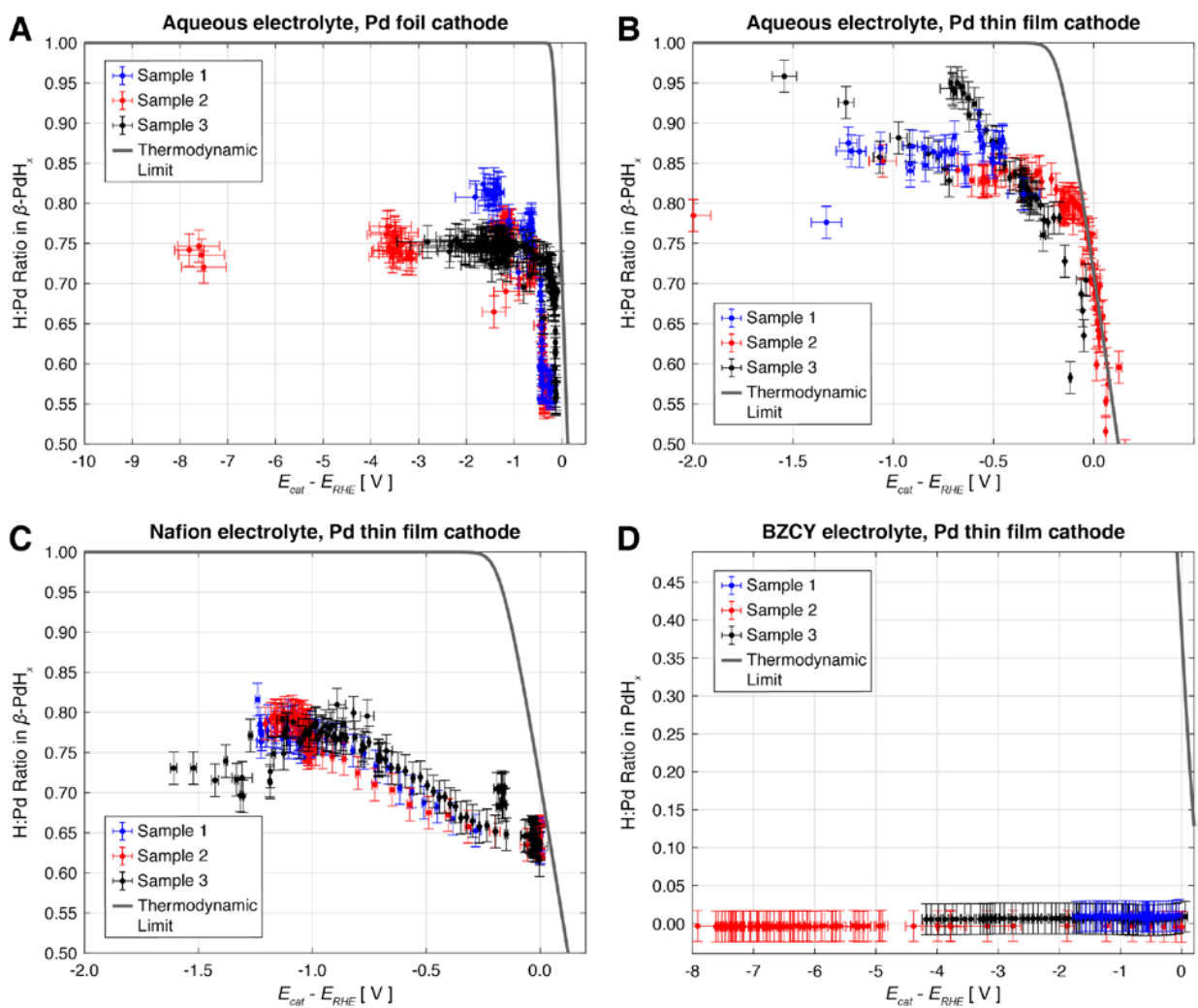

**Figure 6.** PdH$_x$ composition as a function of cathode potential (relative to RHE) for three samples each in the four electrochemical cell configurations tested. The continuous curve represents the thermodynamic limit as determined by the Nernst equation (eqs S7-S9 in the Supporting Information). All results are measured at room temperature except for the BZCY-based cell, which is measured at 400°C. In the aqueous electrolyte cells (A, B), the cathode composition closely follows the thermodynamic limit at small negative cathode potentials, especially in the case of the thin film cathodes. At more negative cathode potentials, the H:Pd ratio plateaus at a value well below the thermodynamically predicted concentration. The thin film cathodes reached a higher maximum H:Pd ratio (0.96) than did the foil cathodes (0.83). (C) Nafion™ thin film cathodes reached a maximum H:Pd ratio of 0.81 at a cathode potential of approximately -1 V before dropping at more negative potentials. (D) BZCY thin film cathodes equilibrated at 400°C did not show any detectable electrochemical H insertion within the range of potentials tested.

Figure 6 distills the time series data from Figures 4 and 5 into plots of H:Pd ratio vs cathode potential (relative to RHE). Each plot includes results for three samples of the same electrochemical cell construction. Also plotted in each instance is the equilibrium H:Pd ratio predicted by the Nernst equation (eqs S7-S9 in the Supporting Information). While these results also show a reproducible relationship when plotted as H:Pd ratio vs current density (see Supporting Information Figure S2), the behavioral trends are more clearly observed when plotted as in Figure 6, due to the stronger dependence of the H:Pd ratio on the cathode potential.

Figures 4, 5, and 6 show that for the aqueous foil, aqueous thin film, and Nafion™ thin film cells, applying a current to the cathode causes significant electrochemical H insertion, leading to the formation of $\beta$-PdH$_x$. In each instance, as the current density is stepwise increased, the corresponding H:Pd values also increase, but the specific behavior differs in each case. In no instance did we observe H:Pd ratios that reached the theoretical Nernstian value of 1.0 with increasing cathode potential; in some instances a maximum in H:Pd is reached, while further ramping of the current density caused a decrease in loading. The aqueous thin film samples showed the highest H:Pd ratios (Fig. 6B), and one sample reached H:Pd = 0.96, the highest value observed in this study. The Nafion™-based cells reached H:Pd = 0.81, which is still significantly above the room-temperature, 1 atm H$_2$ equilibrated value of 0.68. In the BZCY-based cells, no H insertion above background was observed by XRD, even though the measured cathode potential implies a high H:Pd value. At the measurement temperature of 400°C for these cells, the equilibrium H:Pd at 1 atm H$_2$ is <0.01.[31]

**Competing Reaction Mechanisms Determine Steady-State H:Pd Ratio**

In order to understand these results, competing mechanisms of hydrogen insertion and evolution must be taken into account, as shown schematically in Figure 7. In each of the four cell types, a negative cathode potential provides the thermodynamic driving force for H insertion. The electron current flowing through the electrochemical cell contributes to two reactions, H insertion (Reaction 1), which increases H:Pd, and the hydrogen evolution reaction (HER, Reaction 2), which does not. In addition, the atomic H inserted into Pd can diffuse and eventually desorb from the lattice as $H_2$ (Reaction 3), lowering H:Pd. This desorption process does not require charge-transfer and therefore does not contribute to current passing through the cell. The steady-state H:Pd ratio reached in the Pd depends on the relative rates of these three reactions.

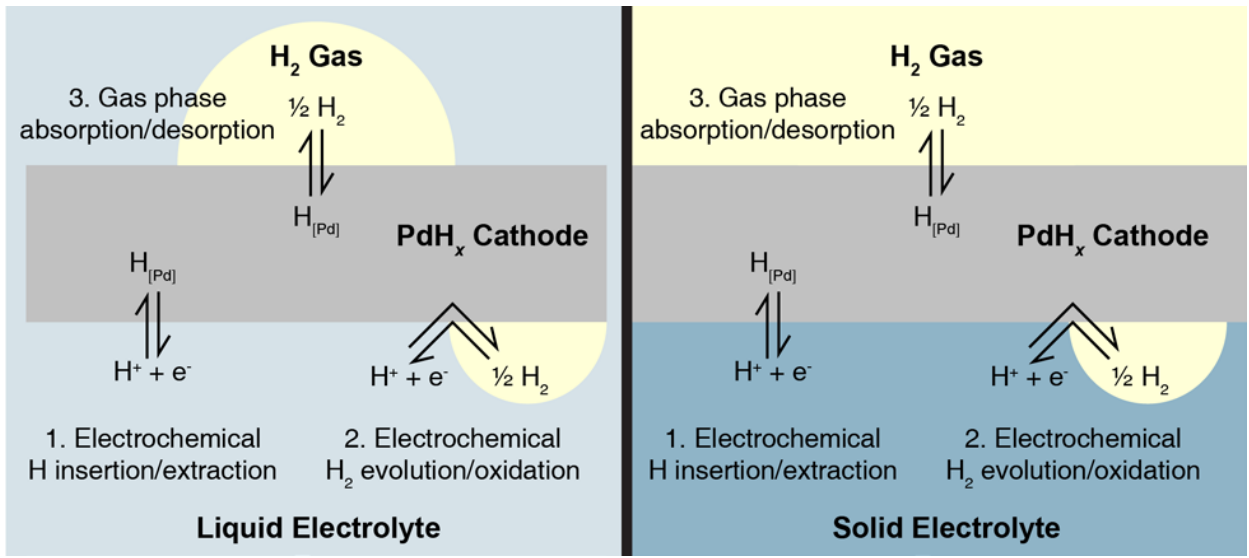

**Figure 7.** Hydrogen reaction processes occurring at the $PdH_x$ electrode/electrolyte interface during electrochemical H insertion or extraction. In the insertion process, $H^+$ in the electrolyte can recombine with electrons to form either (1) H inserted in $PdH_x$ or (2) $H_2$ gas directly. Additionally, (3) H inserted in $PdH_x$ can be released from the cathode to form $H_2$ gas or vice versa, depending on the relative chemical potentials of H in $PdH_x$ and H in the surrounding $H_2$ gas.

**Thin Film Pd Cathodes Yield Higher H:Pd Ratios**

The aqueous electrochemical cell results in Figure 6 show that the thin film Pd cathodes as a group reach higher H:Pd values than do the Pd foil cathodes. The maximum H:Pd ratio reached by a thin film cathode is $0.96 \pm 0.02$ compared to $0.83 \pm 0.02$ for the Pd foil cathode. In addition to the higher absolute value H:Pd ratio, the ramp rate to the steady-state concentration, at the same current density, is much faster for the thin films than the foils. This second effect is to be expected from the the comparative fluxes of hydrogen relative to palladium. However, the higher steady-state H:Pd ratio reached in the thin films may also result from a combination of surface orientation and electrochemomechanical effects, as discussed below.

Because the H flux per Pd atom scales inversely with foil thickness at constant current density (here current density is assumed equivalent to electrochemical H insertion per unit area), this flux is a factor of 500 higher for the thin film than for the Pd foil. The results from the aqueous electrolyte cells show a clear correlation between this flux and the rate at which the H:Pd ratio approaches steady-state. Referring first to Figure 4A, during the time interval 6.8 hr to 8.7 hr, the current density is held constant at -3 $mA/cm^2$ for the foil sample. Over this segment of the experiment, the H:Pd ratio increases from 0.56 to 0.65. During the next segment from 8.7 hr to 10.6 hr, where the current density is held constant at a higher value of -10 $mA/cm^2$. the H:Pd ratio rises to 0.74. These time scales are consistent with the following approximate distribution of the incoming current: 10% of the current directed towards the H insertion reaction (Figure 7, Reaction 1), 90% of the current directed towards HER (Figure 7, Reaction 2). Based on independent measurements of the H desorption rate (Figure 7, Reaction 3), we assume the rate of Reaction 3 to be negligible compared to the other two reactions. For this distribution of the total current, a current density of -3 $mA/cm^2$ should increase the H:Pd ratio by 0.09 in 2.25h. This is in reasonable agreement with the experimental results in Fig. 4A for the Pd foil. And, at higher current densities, the time to reach steady state in the foil sample is reduced, as expected.

In comparison, for the thin film samples, a similar partitioning of the current should result in nearly instantaneous ramping to the steady state H:Pd value, due to the 500-fold lower sample mass. This also is consistent with the results, Figure 4B. (The cause of the decrease in H:Pd at the highest current densities is discussed later.) The minimum current density required to form $\beta$-PdH$_x$ is also smaller in the thin film cathodes than in the foils. In both cases, the cathode starts as single-phase $\alpha$-Pd and transitions to $\beta$-PdH$_x$ under application of current. In the Pd foil (Figure 4A), $\beta$-PdH$_x$ is first observed at a current density of -3 mA/cm$^2$. This first-order phase transformation has an associated mechanical deformation of the cathode due to the lattice expansion, which is clearly observed in Supporting Information Video S2. In contrast, the aqueous thin film cell in Figure 4B immediately and completely transforms to $\beta$-PdH$_x$ upon application of the lowest current density of -2 mA/cm$^2$. Experiments at lower current density were conducted on the thin film cathodes as well, with results shown in Supporting Information Figure S1. Here, the onset of $\beta$-PdH$_x$ is first observed at -0.093 mA/cm$^2$ with complete transformation to $\beta$-PdH$_x$ occuring by -0.290 mA/cm$^2$. Thus, the critical current density for formation of the $\beta$-PdH$_x$ is approximately 30 times lower for the thin film than for the foil.

We determined that H diffusion causing a redistribution of H in the $\beta$-PdH$_x$ cathodes is an unlikely alternative explanation for the lower H:Pd ratio observed in the Pd foil. In the aqueous electrolyte cells, X-rays illuminate the cathode from the top, while electrochemical H insertion primarily occurs through the electrode/electrolyte interface at the bottom. The X-ray attenuation length through the cathodes is approximately 4 μm (Supporting Information Section S1).[48] Therefore, the XRD spectra of the 50 nm thin Pd cathodes reflect the H:Pd ratio averaged over the film thickness. In contrast, the XRD spectra of the 25 μm thick Pd cathode are obtained from the portion of the film farthest from the H insertion interface. However, based on the diffusion coefficient for H in $β$-PdH$_x$ of ~$10^{-11}$ m$^2$/s at room temperature,[49,50] the variation in H:Pd ratio across the 25 μm Pd foil cathode is calculated to be <0.01 over the time duration of the experiment (details provided in Supporting Information Section S4). Therefore, the X-ray observed H:Pd ratio in the foils is expected to reflect a homogenized bulk value.

Another factor that may influence the ramp rate and maximum value of H:Pd is the surface crystallographic orientation of the Pd. As shown in Figure 2A, the sputtered thin films have a <111> preferred orientation. In contrast, the surface texture of the Pd foils shows a slight <220> preferred orientation (Supporting Information Figure S3A). Li *et al.*[51] recently demonstrated that the {100} facets of Pd have higher HER activity than the {111} facets. Additionally, Johnson *et al.*[52] found that the phase transition of PdH$_x$ between $β$ and $α$ phases is slower for nanoparticles with a <111> surface orientation (i.e., octahedral morphology) than for those with <100> surface orientation (cubic morphology). Slower HER and/or gas phase H desorption kinetics at {111} facets could potentially account for the higher H:Pd ratios observed in the <111> oriented Pd thin films. However, since these processes share some fundamental mechanistic steps with electrochemical hydrogen insertion, the rates of this process are likely also affected by the crystallographic orientation, so further study will be necessary to conclusively determine the effect of crystallographic orientation on the achievable H:Pd ratio. Strain effects could also contribute to the higher H content observed in thin films by promoting the H insertion reaction, leading to a higher portion of the incoming current directed towards insertion over HER.[51]

**Electrochemomechanical Damage Affects the Achievable H:Pd Ratio**

When the same 50 nm Pd cathode is used in the aqueous electrochemical cell (Figure 4B) and the Nafion™ cell (Figure 5A), the former reaches a higher H:Pd ratio (0.96 vs 0.81). One explanation for this effect is that the liquid electrolyte - solid Pd interface in the aqueous cell is obviously conformal, as well as being "self-repairing" if the PdH$_x$ morphology changes, due to good wetting by the liquid electrolyte. In contrast, the Nafion™ and BZCY cells have solid electrolyte - solid Pd interfaces which mechanically constrain the Pd film to varying degrees. Note that despite the conformal liquid-solid interface in the aqueous electrolyte cells, the greatest sample-to-sample variation in H:Pd ratio was observed in these cells. We traced this effect to the appearance of H$_2$ gas bubbles in the aqueous electrolyte that block the interface to varying degrees. In contrast, the Nafion™ cells, which only have solid-solid and solid-gas interfaces, show a smaller H:Pd variability.

Electrochemomechanical damage to the Pd cathodes is another important effect that is observed in some form in each of the cell configurations. In the aqueous electrolyte cell, macroscopic mechanical damage can be seen in real-time video of the Pd cathode during electrochemical H insertion (Supporting Information Video S3). As the magnitude of the current increases, the cathode cracks and delaminates from the Kapton window, eventually resulting in the cell failing catastrophically. As shown in Figure 4B, during scans 48 and 49 (15.2 and 15.5 hr), cell failure causes the current density to drop from -40 to -0.01 mA/cm$^2$ and the cathode potential to reach the potentiostat limit set at -11 V. Note that in Figures 4 and 5, for the aqueous electrolyte and Nafion™ cells respectively, as the current is increased beyond a certain point, the H:Pd ratio decreases. We attribute this decrease to cathode damage or delamination.

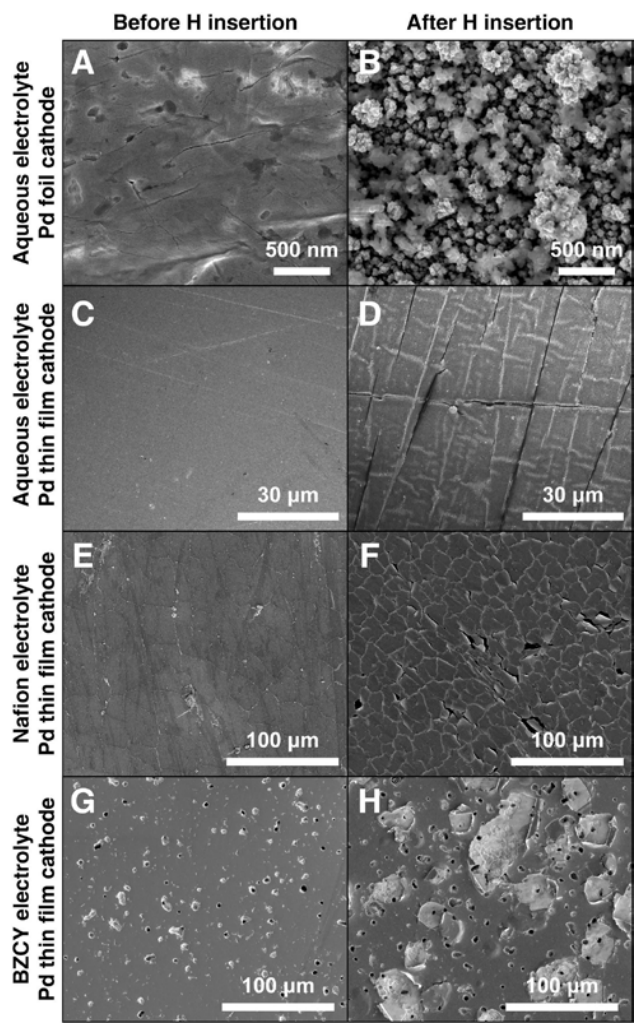

**Figure 8.** Scanning electron micrographs of cathode surface morphology before and after electrochemical H insertion for both foil and thin film cathodes used in aqueous, Nafion™, and BZCY electrolyte cells. After H insertion, nanometer-scale surface roughening was found in aqueous foil cathodes, while micron-scale cracking, spalling, and delamination were observed in thin films utilizing both aqueous and solid electrolytes. Electrochemical H insertion for each sample involved maximum current densities of (B) -300, (D) -40, (F) -105, and (H) -11 mA/cm$^2$.

Observed microscopically, clear electrochemomechanical damage appeared in each of the sample types. Figure 8 shows scanning electron micrographs of a cathode in each of the four electrochemical cell configurations, before and after electrochemical H insertion. Additional images at a variety of magnifications are shown in Supporting Information Figures S6, S7, S8, and S9. The Pd foil cathode surface is severely roughened and comminuted into particulates as a result of hydrogen insertion, Figs. 8 A and B. The thin film Pd cathodes are severely damaged after H insertion, regardless of electrolyte, as seen in Fig. 8 C-H. For the aqueous and Nafion™ electrolytes, the expansion-contraction stresses create widespread cracking leaving domains on the length scale of micrometers. In the BZCY cells, where the solid electrolyte is the least compliant of those studied, even more dramatic electrochemomechanical damage was observed. As Figs. 8 G and H show, after electrochemical insertion of hydrogen, entire grains of solid electrolyte in the sintered BZCY have been ejected, presumably due to the high pressures associated with hydrogen gas evolution. This form of damage was observed repeatedly in experiments on this cell configuration.

**H:Pd Ratio Decreases with Increasing Temperature**

In the BZCY cells, despite the clear evidence for a substantial hydrogen flux as seen by the electrochemomechanical damage, negligible H:Pd ratios are observed in the *operando* XRD experiments (Fig. 5 B and 6 D). The experiments shown are conducted at 400 °C in order to lower the BZCY electrolyte resistance. (Others, not shown, were carried out over a wide range of temperatures from 210 to 750 °C, with similar results.) At this temperature, the equilibrium H:Pd ratio is nearly zero at 1 atm $H_2$. We believe that the HER reaction (Figure 7, Reaction 2) and/or the gas phase $H_2$ desorption reaction (Figure 7, Reaction 3) is simply too rapid at 400 °C to permit significant bulk hydrogen to be obtained.

Finally, the results above emphasize why *operando* structural measurements are preferable over purely electrical or electrochemical techniques. The trends in the H:Pd ratio observed in our measurements could easily be misinterpreted if only the resistance ratio ($R/R_0$) is used to characterize H:Pd ratio, as described in Supporting Information Section S6. In the BZCY cells, we also attempted to use the open-circuit (Nernst) potential to measure the instantaneous H:Pd ratio. During the experiment shown in Figure 5B, the Nernst potential was measured for 1 second every 5 minutes. As described in Supporting Information Section S3, these measurements indicated high H:Pd ratios in some experiments, but the XRD measurements show negligible hydrogen retention.

We provide an extensive Supporting Information Section with additional details on many of the analyses discussed. In addition to material already referred to, Section S5 examines the use of the Nafion™-based solid electrolyte cells as a vehicle for investigating surface reaction kinetics. Pd and $PdH_x$ are potentially useful electrocatalysts for reactions including $CO_2$ reduction and the electrochemical hydrogen evolution reaction (HER). Electrochemical H insertion provides a potential strategy for modulating catalytic activity in this material system. The lattice expansion and composition change associated with increasing H:Pd ratio can affect the binding energy of surface-adsorbed catalytic intermediates, and therefore affect the reaction kinetics. At present, these effects are difficult to understand and control, because prior to this study, it has been challenging to accurately measure the lattice parameter and H:Pd ratio of $β$-$PdH_x$ during electrochemistry. Amongst the cell types we developed, the Nafion™-based cells are advantageous for such studies due to low, uniform and well-characterized series resistance and the absence of interference from gas bubbles. In S5, we obtain preliminary results showing the dependence of HER kinetic parameters on overpotential and H:Pd ratio.

## Conclusions

The electrochemical insertion of hydrogen into palladium via liquid and solid state electrolytes has been investigated. We show that the H:Pd ratio achieved is the result of a dynamic balance between the rate of hydrogen insertion and evolution from the Pd lattice, and use *operando* X-ray diffraction to characterize this process. At room temperature and ambient pressure, H:Pd ratios as high as 0.96 ± 0.02 are obtained. Higher H:Pd ratios are observed in thin films than in thicker foils, at least in part due to a higher hydrogen flux per palladium. Higher H:Pd ratios are reached using liquid than solid electrolytes, due to their conformal, self-healing nature. In all electrolytes, electrochemomechanical damage resulting from the large volume change induced upon hydrogen insertion strongly affects the achievable H:Pd ratio.

The results highlight the difficulty of reaching stoichiometric limiting compositions such as PdH. At H:Pd ratios close to 1, a large overpotential is required to insert hydrogen rapidly enough to compensate for the enormous driving force for $H_2$ desorption from the Pd lattice. Concurrently, HER is a facile competing reaction to H insertion which, even at modest applied potentials, diverts much of the applied current towards gas evolution. Simultaneously, electrochemomechanical damage to the palladium increases the available surface area for hydrogen evolution and decreases the extent of electrolyte-palladium contact.

Nonetheless, the understanding embodied in these results point to new design rules for achieving high hydrogen concentrations in metal hydrides, and suggest several possible pathways to overcoming the challenges. Reducing the electrochemical $H_2$ evolution reaction rate through surface modifiers,[53] or by optimizing for palladium surface orientation effects,[51,52] could increase the achievable H:Pd ratio. Designing thin cathode structures that are resistant to electrochemomechanical damage is another route, including very thin foils or folding morphologies that allow volume expansion and contraction without mechanical failure.

## Acknowledgements

The authors thank Dr. Charles Settens for assistance with XRD measurements and data analysis. The authors also thank Dr. David Fork, Dr. Ross Koningstein, and Matthew Trevithick for helpful discussions about data analysis and interpretation. XRD and SEM measurements were performed at the MIT Center for Materials Science and Engineering, a MRSEC Shared Experimental Facility supported by the National Science Foundation under award number DMR-14-19807. Financial support was provided by Google LLC. DY acknowledges support from a National Science Foundation Graduate Research Fellowship under Grant No. 1122374.

## Supporting Information

The Supporting Information is available free of charge online at

https://docs.google.com/document/d/1OTR70IUz3OWQZvx3UFzGsSmudrd4F8XLFj1Vi_msN8I/edit#

Details of procedure used to create correlation between $β$-PdH$_x$ lattice parameter and H:Pd ratio, calculations about H concentration gradient due to diffusion inside $β$-PdH$_x$ cathodes, electrochemical measurements of H:Pd ratio, X-ray attenuation length in $β$-PdH$_x$ cathodes, electrochemical hydrogen evolution reaction kinetics, representative raw XRD spectra from each electrochemical cell configuration, additional scanning electron microscope images, energy dispersive X-ray spectroscopy elemental composition maps, videos of aqueous thin film and foil cathodes during electrochemical H insertion.

# TOC Graphic

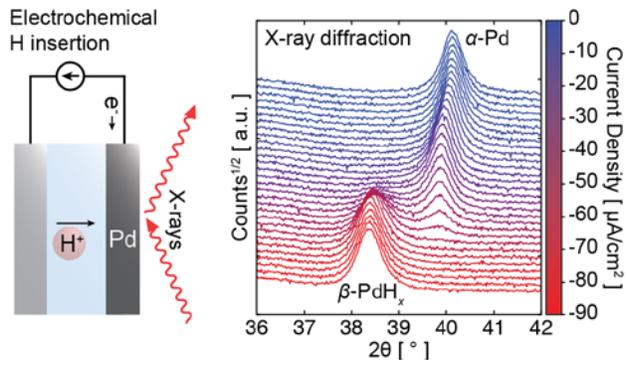